\documentclass{iopconfser}

\usepackage{amsmath}
\usepackage{wrapfig}
\usepackage{graphicx}

\begin{document}

\title{Testing Gravitation from Light-second to Cosmological Scales with Radio Pulsars}

\author{Emmanuel Fonseca$^{1,2}$}

\affil{$^1$Dept. of Physics and Astronomy, West Virginia University, Morgantown, WV, USA}

\affil{$^2$Center for Gravitational Waves and Cosmology, West Virginia University, Morgantown, WV, USA}

\email{emmanuel.fonseca@mail.wvu.edu}

\begin{abstract}
Pulsars are spinning neutron stars typically observed as pulses emitted at radio wavelengths. These pulsations exhibit a rotational stability that rival the best atomic clocks, making pulsars one of the most important tools for resolving gravitational phenomena in extreme environments. I will present an overview of the ways in which radio pulsars can be used to test strong-field gravity and observe gravitational radiation, both in the context of historical and ongoing experiments. I will also describe how these measurements can be translated to sought-after quantities like the masses and moments of inertia of neutron stars.
\end{abstract}

\section{Introduction}
The study of gravitation relies on the availability of ``rulers" and ``clocks" to characterize spacetime. In this sense, the discovery of the first radio pulsar by J. Bell-Burnell was a fortuitous moment in gravitational research due to the clock-like stability immediately evident in its pulsations \cite{hbp+68}. Within its first decade, pulsar science saw the discovery of new sources and common properties (e.g., \cite{hmt75}) that have since established pulsars as natural, high-precision ``cosmic clocks" that grant access to spacetime environments beyond the Solar System. The implications from pulsars' clock-like behavior were quickly recognized as being useful for the study of A. Einstein's general relativity (GR), though such ideas were proposed at a time when observational possibilities were not yet realized (e.g., \cite{hof68}). As described below, these possibilities gradually changed into opportunities through the discovery of new kinds of sources and the development of rigorous methods with which to study the timing properties of pulsars.

For this contribution, I summarize the landscape of testing and probing the consequences of GR using radio pulsars. This summary is intentionally more brief than those found in academic literature, and the interested Reader is encouraged to seek out these excellent reviews (e.g., \cite{sta03,lor05,fw24,hu25}). I partition the landscape into two categories due to the scales at with GR can be studied. In Section \ref{sec:binaries}, I overview the ways in which binary pulsars are used to study the relativistic two-body problem at $O$(light-second) distances between components. In Section \ref{sec:ptas}, I overview how ongoing ``pulsar timing array" (PTA) experiments are likely to confirm the existence of gravitational waves (GWs) arising from a cosmological population of high-mass, inspiraling black holes in the near future. I provide a summary of the key and foundational method -- {\it pulsar timing} -- that enables both applications in the following subsection, so that any Reader can become familiar with its power and underlying assumptions.

\subsection{The Key Method: Pulsar Timing}
\label{subsec:timing}

As the ``gateway" observing method, pulsar timing consists of analyzing and interpreting variations in pulse times of arrival (TOAs) as recorded by telescopes on Earth \cite{lk12}. TOAs encode physical information that arise from intrinsic spin variations of the pulsar, electromagnetic effects from signal propagation through ionized media, and the non-inertial nature of the pulsar and/or an observer on Earth. The ultimate product from pulsar timing is a ``timing model" that parameterizes all relevant effects into explicit, additive terms that each correspond to a specific phenomenon. 

The art of pulsar timing fundamentally relies on two assumptions:

\begin{enumerate}
    \item pulses occur over time intervals that reflect the spin frequency of the pulsar ($\nu_{\rm s}$), which is constant over short timescales and can vary slowly over time;

    \item the time-averaged shape of the pulse remains constant over time.
\end{enumerate}

\noindent When applied to a non-accelerated pulsar, the first assumption allows for the definition of a phase of pulsar rotation ($\phi$) as a function of time ($t$) using a Taylor expansion, i.e., $\phi(t) = \phi_0 + \nu_{\rm s}(t-t_0) + \dot{\nu}_{\rm s}(t-t_0)^2$ + ..., where a dot refers to a time derivative and $\phi_0$ is the phase at a reference time $t_0$. The second assumption allows for an unambiguous association of any TOA (and thus $\phi_0$) with a specific pulse feature. The effort to measure TOAs therefore becomes an iterative, template-matching process which allows for data-driven (and thus model-independent) optimization of TOA accuracy and precision \cite{tay93}. In practice, TOA precision of $O(0.1-1\textrm{ }\mu\textrm{s})$ is achievable for a small but important segment of the known pulsar population.

When taken together, the two assumptions above enable pulsar timing to be a high-accuracy method with very little model dependency. The above framework can be generalized for a non-inertial observer and an accelerated pulsar by treating $t$ as a proper time; the TOA is then treated as observer's coordinate time ($t'$) that must be corrected in order to reflect a time measured in an inertial reference frame. These corrections correspond to the physical phenomena affecting the pulsar and/or observer, and the collection of their parameters define the timing model.

\section{The Light-second Scale: Pulsars in Orbital Systems}
\label{sec:binaries}

The power of pulsar timing is especially apparent for pulsars in orbital systems, where timing variations due to orbital motion exceed the timing precision of individual TOAs by many orders of magnitude. Indeed, the most consequential discovery of a binary pulsar is arguably the very first: in 1974, structured modulations in a newly discovered pulsar were shown to arise from Doppler shifts due to orbital motion in a compact and dynamically ``clean" binary system \cite{ht75}; within several months, it was pointed out that the system was sufficiently relativistic to yield an unambiguous test for the existence of GWs through the gradual inspiral of the system, in reaction to the corresponding loss of orbital energy \cite{wag75}; several high-precision ``timing models" were developed to optimally characterize all aspects of relativistic information encoded in binary-pulsar TOAs (e.g., \cite{dt92}); and continued observations of this binary pulsar led to the first measurement of orbital decay, at a rate entirely consistent with the emission of GWs as predicted by GR \cite{tw89}. An historical account regarding the discovery and developments motivated by this source -- now commonly known as the Hulse-Taylor pulsar -- is wonderfully transcribed by \cite{dam15}.

The Hulse-Taylor pulsar was discovered fifty years prior to the submission of this proceedings. Since this discovery, the population of known orbital pulsars has grown to include a diverse range of gravitational environments. For example, the Hulse-Taylor pulsar was the first of $\sim$ 20 ``double-neutron-star" (DNS) systems now known as of submission of this contribution. By contrast, the subpopulation of pulsars orbiting white dwarfs (WDs) is a factor of $\sim 10$ larger. The relative sizes and orbital properties of these subpopulations are understood to reflect the different formation channels within binary stellar evolution \cite{tv23}. However, a number of ``outlier" systems tend to exist in globular clusters, where many-body interactions are believed to have influenced the orbital evolution of the pulsars they harbor (e.g., \cite{fre13}). Both DNS and pulsar-WD systems nonetheless provide novel avenues for testing different aspects of gravitation in the strong-field regime.

\begin{wrapfigure}{r}{0.5\textwidth}
    \centering
    \includegraphics[width=0.5\textwidth]{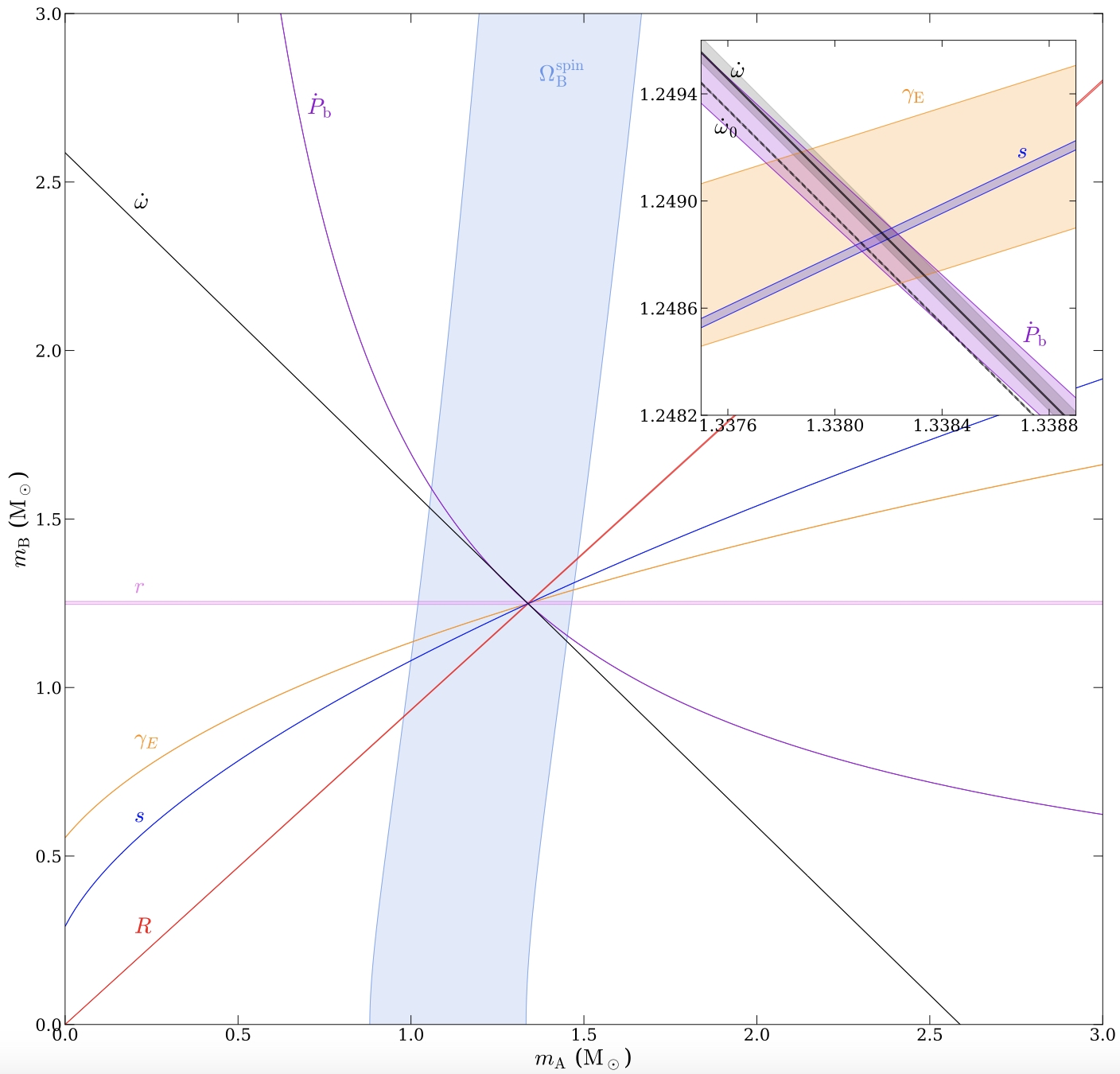}
    \caption{PK effects observed in the double-pulsar binary system, each represented as a set of shaded curves. The extent of each shaded region corresponds to the 68.3\% confidence region. All curves intersect at a common region as shown in the inset; their union illustrates the self-consistency of GR in explaining these phenomena. This figure is reproduced from \cite{ksm+21} under Creative Commons BY 4.0.}
    \label{fig:m1m2}
\end{wrapfigure}

DNS pulsars provide the ideal conditions for testing the relativistic two-body problem through the observation of ``post-Keplerian" (PK) effects, These PK effects -- including orbital decay, time dilation, and apsidal motion -- can be measured through pulsar timing in a theory-independent manner, allowing tests for self-consistency of any viable gravitational theory \cite{dt92}. The hallmark DNS system to date is the ``double-pulsar" binary system due to its compactness \cite{bdp+03} and the unique circumstance of both neutron stars being observed as pulsars for several years \cite{lbk+04}.\footnote{The spin angular momentum vector of the slow-spinning pulsar in this system, PSR J0737$-$3039B, was confirmed to undergo de Sitter precession \cite{bkk+08}. This pulsar precessed out of our line of sight around 2008, but is believed to precess back into view sometime within the next decade \cite{pmk+10}.} Long-term timing of the fast-spinning pulsar in this system, PSR J0737$-$3039A, has yielded the most robust tests of GR in the strong-field regime with measurements of 7 PK effects as shown in Figure \ref{fig:m1m2} \cite{ksm+21}. These tests include the field-leading test of orbital decay due to GW emission, which agrees with GR at the 0.01\% level. Moreover, PSR J0737$-$3039A/B is also the ideal DNS system for testing next-to-leading-order corrections of photon propagation in spacetime curvature due to its extremely high degree of orbital inclination \cite{hkc+22}.

Despite its vast reach, pulsar timing is not the only means with which to observe and measure PK effects in DNS systems. For example, the Hulse-Taylor system gradually exhibits slow, secular changes in its pulse profile at a rate consistent with de Sitter precession of the spin angular momentum vector ${\bf S}_{\rm p}$ \cite{wrt89}. Secular changes can also be observed as changes in angles that define the sense of ${\bf S}_{\rm p}$, which can be measured through the modeling of polarization ``swings" as a function of rotational phase $\phi$ (e.g., \cite{cwb90}). The union of measurements from profile and polarization evolution can be used to constrain orbital geometry \cite{kra98} or to directly measure the rate of de Sitter precession \cite{sta04,dkl+19}. The double-pulsar system offers an entirely different way to measure the de Sitter precession rate for this system: the pulsations from J0737$-$3039A are regularly eclipsed by the magnetic field of J0737$-$3039B, and evolution in the eclipsing signal matches the predictions of orientation changes due to de Sitter precession \cite{bkk+08}.

The dynamics of pulsar-WD binary systems can also be described as the interaction of point particles, and thus yield PK effects over time (e.g., \cite{fpe+16}). However, most pulsar-WD systems are not sufficiently compact or eccentric to yield measurements of time dilation and apsidal motion. Therefore, these systems yield a smaller number of opportunities to test GR. Nevertheless, pulsar-WD systems offer different and intriguing avenues for testing gravitation on light-second scales. For example, the first evidence of Lense-Thirring precession of a binary pulsar was found in a pulsar-WD system \cite{kbv+20}. Moreover, many alternative theories that deviate from GR rely on the violation of the strong equivalence principle (SEP). Violation of the SEP introduces a dependence of orbital motion on the composition of the self-gravitating bodies, which manifests most strongly in binary systems where the components have different mass densities. This condition is most naturally met in compact pulsar-WD binary systems. Indeed, such systems have been used to place limits on the presence of dipolar gravitational waves that are predicted by tensor-scalar theories of gravity, as well as temporal variations in the gravitational constant $G$ \cite{fwe+12,zsd+15}. SEP violation has itself been stringently tested using a radio pulsar in a hierarchical triple system with two WDs \cite{agh+18,vcf+20}. 

The confirmation of GR with binary pulsars provides the means to accurately measure neutron-star masses with high precision through pulsar timing \cite{of16}. These mass measurements are prized in various communities for the constraining power they offer to different branches of active research. For example, the growing distribution of pulsar masses provides clues into the delineation between formation channels and supernova mechanisms (e.g., \cite{spr10}). In pulsar-WD systems, measurements of WD masses are useful for probing the expected correlation between orbital parameters that reflect prolonged mass transfer during the ``spin-up" phase in the progenitor system \cite{ts99}. The most sought-after measurements tend to be of pulsars with the highest masses (e.g., \cite{cfr+20}), as they can be used to effectively rule out proposed equations of state that depend on ``exotic" states of matter that cannot be reproduced through terrestrial experiment. Even in the multimessenger era, pulsar mass measurements remain a key and trusted data set that is made possible through rigorous testing of GR.

\section{The Cosmological Scale: GWs and Pulsar Timing Arrays}
\label{sec:ptas}

The notion that pulsars could be used to directly detect GWs was proposed within the first decade of pulsar astronomy \cite{saz78,det79}. These early works postulated that binary systems containing objects with masses of $O(10^9$ $M_\odot$), orbiting with separations of $O(1\textrm{ AU})$, could be a natural source of GWs with nanohertz (nHz) frequencies; these spacetime-metric fluctuations would then impart $O(0.1\textrm{ }\mu$s) variations in pulsar TOAs. It is worth noting that these works were published in an era where supermassive black holes (SMBHs) -- the only astrophysical objects that could realize the conditions stated above -- were largely considered to be hypothetical objects of unclear origins. However, this era also yielded a growing number of quasar observations and the discovery of Sagittarius A* \cite{bb74}, all of which implied the existence of SMBHs (e.g., \cite{sal64}). Therefore, the most regular and rapidly rotating pulsars in the known population were quickly understood to be useful for placing meaningful limits on nHz-frequency GWs originating from a population of SMBH binary systems.

\begin{figure}
    \centering
    \includegraphics[width=1.0\linewidth]{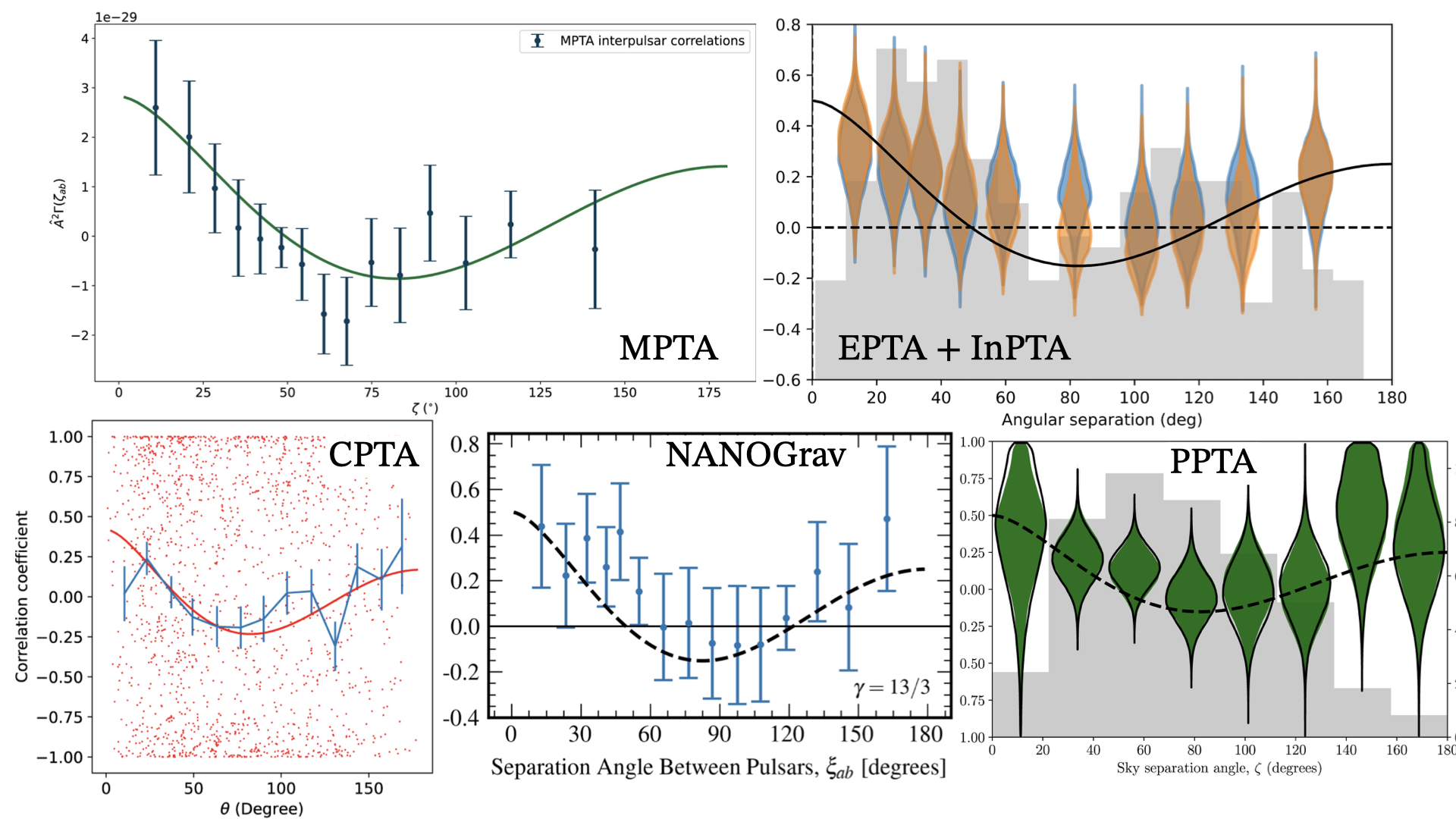}
    \caption{Spatial correlations between pairs of MSPs observed by various PTAs. In each panel, the solid- or dashed-line curves represent the expected HD correlation due to a stochastic GW background. Each panel is adapted from recent works produced by: the Chinese PTA (CPTA; \cite{xcg+23}); the European and Indian PTAs (EPTA + InPTA; \cite{aaa+23a}); the MeerKAT PTA (MPTA; \cite{msr+25}); the North American Nanohertz Observatory for Gravitational Waves (NANOGrav; \cite{aaa+23a}); and the Parkes PTA (PPTA; \cite{rzs+23}). The CPTA panel is reproduced with permission granted by the publishing journal; all other panels are reproduced from the cited works under Creative Commons BY 4.0.}
    \label{fig:hd}
\end{figure}

The discovery of the first millisecond pulsar (MSP) demonstrated that such regular pulsars indeed exist \cite{bkh+82}, proving that the direct detection of nHz-frequency GWs was a realistic endeavor. However, additional MSP discoveries and dedicated timing observations showed that no MSP is a perfect clock: all pulsars exhibit their own degrees of stochastic noise processes as reflected in their TOA residuals \cite{sc10}. These sources of statistical noise arise from several mechanisms, such as imperfectly modeled variations in pulse dispersion or magnetospheric torques that affect pulsar rotation. The cosmological GW signal itself is expected to be statistically ``red", representing a superposition of GWs. The use of a single pulsar is therefore not robust for directly detecting GWs.

A further complication is that metric fluctuations due to nHz-frequency GWs are expected to arise both at the source of pulsed radiation (i.e., the pulsar) and the observing point on Earth \cite{ew75}. The GW signal can therefore be partitioned into ``pulsar" and ``Earth" terms, respectively (e.g., \cite{lwk+11}). However, the Earth term is present in all Earth-based TOA measurements and is therefore a ``correlated" signal, i.e., a red-noise process that is common in all pulsar TOA residuals. This expected correlation led to the development of a PTA prototype, where TOA-residual data for an ensemble of four MSPs were cross-correlated to place limits on the amplitude of a common red-noise process present in the ensemble \cite{hd83}. The first PTA prototype showed that the correlation between pairs of MSPs follows a predictable trend, as a function of angular separation between pulsars, that is specific to the GW polarization states predicted by GR. In pulsar literature, this trend is referred to as the ``Hellings-Downs" (HD) correlation.

Since this initial effort, various teams of pulsar scientists have organized to form large-scale PTAs around the Earth. These PTAs currently span five continents and are distinguished by a specific set of radio observatories that can observe specific MSPs across parts of the celestial sphere that are accessible to each telescope. While each PTA operates independently, their union -- the International Pulsar Timing Array (IPTA; e.g., \cite{pdd+19}) -- is an active and growing collaboration that combines timing data for placing the strongest constraints on the stochastic GW background.

PTAs fundamentally rely on the pulsar-timing method to form TOA residuals for cross-correlation. This aspect of GW detection with PTAs therefore requires the meticulous study of all individual MSPs and their noise properties in order to establish them as ``well-characterized detectors." This fact has borne a considerable amount of ancillary science in pulsar astrophysics, including (but not limited to) the discovery of the highest mass neutron star currently \cite{cfr+20} known and the subsequent efforts to measure its radius (eg., \cite{rwr+21}). Moreover, this synergy between GW and PTA science has motivated the development of new pulsar-timing methods designed to extract as much accuracy and precision from pulsars as possible. 

Various PTAs released the first evidence of a stochastic GW background in the nHz frequency band within the last few years \cite{xcg+23,aaa+23b,aaa+23a,rzs+23,msr+25}. Constraints on the HD correlation from these efforts is shown in Figure \ref{fig:hd}. The claims are clear to specifiy ``evidence" for, and not ``detections of", HD correlation due to not (yet) achieving significance thresholds as established by PTA partners \cite{adg+23}. However, these constraints are sufficiently powerful to allow for searches of anisotropy in the background (e.g., \cite{gnt+25}), as well as deviations in background statistics that could be associated with ``new physics", i.e., non-SMBH sources of the stochastic GW background such as first-order cosmological phase transitions, domain walls, and cosmic strings (e.g., \cite{aaa+23c,wch24}). Ongoing expansion of PTA data sets will lead with a significant detection of the stochastic GW background in the coming years. One of the major, ongoing efforts is to combine and jointly analyze all PTA data for placing the strongest constraints on HD correlation using all available data. The results of this global effort are expected to be publsished within the next 1--2 years.

\section{Future Prospects}

Nearing its 60th year, pulsar science remains a vibrant field that is propelled by its sensitivity to gravitational physics. Part of its continued relevance is the inference that many thousands of pulsar systems will become discoverable as more sensitive telescopes are built and used for pulsar science: studies in population synthesis suggest that $O(10^5)$ pulsars should be observable from Earth while only $O(10^3)$ are known \cite{lpr+19}, indicating that the ``discovery space" in pulsar science remains largely unexplored. Additional discoveries are therefore likely to provide novel opportunities to test GR, whether they are usable for per-system (i.e., light-second) or PTA (i.e., cosmological) experiments.

While many possibilities may come to fruition, several milestones are within reach:

\begin{itemize}
    \item {\it The light-second scale}: continued observations of compact DNS systems will yield measurements of new PK phenomena and thus yield new tests of GR, such as the first-order deformation of the orbit from purely elliptical motion (e.g., \cite{wh16}); in several case, it will likely become possible to separately measure the timing signature associated with Lense-Thirring precession, which yields another test of GR and the moment of inertia of the pulsar (e.g., \cite{hf24}).
    
    \item {\it The cosmological scale}: beyond the GW background, PTAs are working towards building their sensitivty to resolvable (or ``continuous wave") sources, along with transient events such as ``bursts with memory" (e.g., \cite{mcc14}); all PTAs are growing the number of MSPs that they observe, which is one of the key conditions for reliably detecting HD correlations and accessing new tests of GR (e.g., \cite{wil98}); 
\end{itemize}

\noindent We therefore expect many exciting moments ahead, made possible through collaborative spirit, and we encourage all interested readers to join us. \\

\noindent {\bf Acknowledgements}:
I am grateful to the anonymous referee for their constructive feedback on this manuscript, and to the organizers of the GR24-Amaldi16 conference for their invitation to deliver this talk. I am supported by the National Science
Foundation under grant AST-2407399.

\newpage

\bibliography{iopconfser-template.bbl}

\begin{thebibliography}{10}

\bibitem{hbp+68}
{Hewish} A, {Bell} SJ, {Pilkington} JDH, {Scott} PF, {Collins} RA.
\newblock {Observation of a Rapidly Pulsating Radio Source}.
\newblock {\it Nature}. 1968 Feb;217(5130):709--713.

\bibitem{hmt75}
{Helfand} DJ, {Manchester} RN, {Taylor} JH.
\newblock {Observations of pulsar radio emission. III. Stability of integrated
  profiles.}
\newblock {\it ApJ}. 1975 Jun;198:661--670.

\bibitem{hof68}
{Hoffmann} B.
\newblock {Pulsars and a Possible New Test of General Relativity}.
\newblock {\it Nature}. 1968 May;218(5142):667--668.

\bibitem{sta03}
{Stairs} IH.
\newblock {Testing General Relativity with Pulsar Timing}.
\newblock {\it Liv~Rev~Rel}. 2003 Dec;6(1):5.

\bibitem{lor05}
{Lorimer} DR.
\newblock {Binary and Millisecond Pulsars}.
\newblock Liv~Rev~Rel. 2005 Dec;8(1):7.

\bibitem{fw24}
{Freire} PCC, {Wex} N.
\newblock {Gravity experiments with radio pulsars}.
\newblock {\it Liv~Rev~Rel}. 2024 Dec;27(1):5.

\bibitem{hu25}
{Hu} H.
\newblock {Unlocking gravity and gravitational waves with radio pulsars:
  advances and challenges}.
\newblock {\it Astrophys~Space~Sci}. 2025 Jul;370(7):74.

\bibitem{lk12}
{Lorimer} DR, {Kramer} M.
\newblock {Handbook of Pulsar Astronomy}; 2012.

\bibitem{tay93}
{Taylor} JH.
\newblock {Pulsar Timing and Relativistic Gravity}.
\newblock Phil~Trans~Roy~Soc~Lon~Ser~A. 1992 Oct;341(1660):117--134.

\bibitem{ht75}
{Hulse} RA, {Taylor} JH.
\newblock {Discovery of a pulsar in a binary system.}
\newblock ApJL. 1975 Jan;195:L51--L53.

\bibitem{wag75}
{Wagoner} RV.
\newblock {Test for the existence of gravitational radiation.}
\newblock ApJL. 1975 Mar;196:L63--L65.

\bibitem{dt92}
{Damour} T, {Taylor} JH.
\newblock {Strong-field tests of relativistic gravity and binary pulsars}.
\newblock {\it Phys Rev D}. 1992 Mar;45(6):1840--1868.

\bibitem{tw89}
{Taylor} JH, {Weisberg} JM.
\newblock {Further Experimental Tests of Relativistic Gravity Using the Binary
  Pulsar PSR 1913+16}.
\newblock ApJ. 1989 Oct;345:434.

\bibitem{dam15}
{Damour} T.
\newblock {1974: the discovery of the first binary pulsar}.
\newblock {\it Class Quant Grav}. 2015 Jun;32(12):124009.

\bibitem{tv23}
{Tauris} TM, {van den Heuvel} EPJ.
\newblock {Physics of Binary Star Evolution. From Stars to X-ray Binaries and
  Gravitational Wave Sources}; 2023.

\bibitem{fre13}
{Freire} PCC.
\newblock {The pulsar population in Globular Clusters and in the Galaxy}.
\newblock In: {van Leeuwen} J, editor. Neutron Stars and Pulsars: Challenges
  and Opportunities after 80 years. vol. 291 of IAU Symposium; 2013. p.
  243--250.

\bibitem{bdp+03}
{Burgay} M, {D'Amico} N, {Possenti} A, {Manchester} RN, {Lyne} AG, {Joshi} BC,
  et~al.
\newblock {An increased estimate of the merger rate of double neutron stars
  from observations of a highly relativistic system}.
\newblock {\it Nature}. 2003 Dec;426(6966):531--533.

\bibitem{lbk+04}
{Lyne} AG, {Burgay} M, {Kramer} M, {Possenti} A, {Manchester} RN, {Camilo} F,
  et~al.
\newblock {A Double-Pulsar System: A Rare Laboratory for Relativistic Gravity
  and Plasma Physics}.
\newblock {\it Science}. 2004 Feb;303(5661):1153--1157.

\bibitem{bkk+08}
{Breton} RP, {Kaspi} VM, {Kramer} M, {McLaughlin} MA, {Lyutikov} M, {Ransom}
  SM, et~al.
\newblock {Relativistic Spin Precession in the Double Pulsar}.
\newblock {\it Science}. 2008 Jul;321(5885):104.

\bibitem{pmk+10}
{Perera} BBP, {McLaughlin} MA, {Kramer} M, {Stairs} IH, {Ferdman} RD, {Freire}
  PCC, et~al.
\newblock {The Evolution of PSR J0737-3039B and a Model for Relativistic Spin
  Precession}.
\newblock ApJ. 2010 Oct;721(2):1193--1205.

\bibitem{ksm+21}
{Kramer} M, {Stairs} IH, {Manchester} RN, {Wex} N, {Deller} AT, {Coles} WA,
  et~al.
\newblock {Strong-Field Gravity Tests with the Double Pulsar}.
\newblock Phys Rev X. 2021 Oct;11(4):041050.

\bibitem{hkc+22}
{Hu} H, {Kramer} M, {Champion} DJ, {Wex} N, {Parthasarathy} A, {Pennucci} TT,
  et~al.
\newblock {Gravitational signal propagation in the double pulsar studied with
  the MeerKAT telescope}.
\newblock AAp. 2022 Nov;667:A149.

\bibitem{wrt89}
{Weisberg} JM, {Romani} RW, {Taylor} JH.
\newblock {Evidence for Geodetic Spin Precession in the Binary Pulsar PSR
  1913+16}.
\newblock {\it ApJ}. 1989 Dec;347:1030.

\bibitem{cwb90}
{Cordes} JM, {Wasserman} I, {Blaskiewicz} M.
\newblock {Polarization of the Binary Radio Pulsar 1913+16: Constraints on
  Geodetic Precession}.
\newblock {\it ApJ}. 1990 Feb;349:546.

\bibitem{kra98}
{Kramer} M.
\newblock {Determination of the Geometry of the PSR B1913+16 System by Geodetic
  Precession}.
\newblock {\it ApJ}. 1998 Dec;509(2):856--860.

\bibitem{sta04}
{Stairs} IH, {Thorsett} SE, {Arzoumanian} Z.
\newblock {Measurement of Gravitational Spin-Orbit Coupling in a Binary-Pulsar
  System}.
\newblock {\it Phys~Rev~Lett}. 2004 Sep;93(14):141101.

\bibitem{dkl+19}
{Desvignes} G, {Kramer} M, {Lee} K, {van Leeuwen} J, {Stairs} I, {Jessner} A,
  et~al.
\newblock {Radio emission from a pulsar{\textquoteright}s magnetic pole
  revealed by general relativity}.
\newblock {\it Science}. 2019 Sep;365(6457):1013--1017.

\bibitem{fpe+16}
{Fonseca} E, {Pennucci} TT, {Ellis} JA, {Stairs} IH, {Nice} DJ, {Ransom} SM,
  et~al.
\newblock {The NANOGrav Nine-year Data Set: Mass and Geometric Measurements of
  Binary Millisecond Pulsars}.
\newblock {\it ApJ}. 2016 Dec;832(2):167.

\bibitem{kbv+20}
{Krishnan} VV, {Bailes} M, {van Straten} W, {Wex} N, {Freire} PCC, {Keane} EF,
  et~al.
\newblock {Lense-Thirring frame dragging induced by a fast-rotating white dwarf
  in a binary pulsar system}.
\newblock {\it Science}. 2020 Jan;367(6477):577--580.

\bibitem{fwe+12}
{Freire} PCC, {Wex} N, {Esposito-Far{\`e}se} G, {Verbiest} JPW, {Bailes} M,
  {Jacoby} BA, et~al.
\newblock {The relativistic pulsar-white dwarf binary PSR J1738+0333 - II. The
  most stringent test of scalar-tensor gravity}.
\newblock {\it MNRAS}. 2012 Jul;423(4):3328--3343.

\bibitem{zsd+15}
{Zhu} WW, {Stairs} IH, {Demorest} PB, {Nice} DJ, {Ellis} JA, {Ransom} SM,
  et~al.
\newblock {Testing Theories of Gravitation Using 21-Year Timing of Pulsar
  Binary J1713+0747}.
\newblock {\it ApJ}. 2015 Aug;809(1):41.

\bibitem{agh+18}
{Archibald} AM, {Gusinskaia} NV, {Hessels} JWT, {Deller} AT, {Kaplan} DL,
  {Lorimer} DR, et~al.
\newblock {Universality of free fall from the orbital motion of a pulsar in a
  stellar triple system}.
\newblock {\it Nature}. 2018 Jul;559(7712):73--76.

\bibitem{vcf+20}
{Voisin} G, {Cognard} I, {Freire} PCC, {Wex} N, {Guillemot} L, {Desvignes} G,
  et~al.
\newblock {An improved test of the strong equivalence principle with the pulsar
  in a triple star system}.
\newblock {\it Astron~Astrop}. 2020 Jun;638:A24.

\bibitem{of16}
{{\"O}zel} F, {Freire} P.
\newblock {Masses, Radii, and the Equation of State of Neutron Stars}.
\newblock {\it Ann Rev Astron Astrop}. 2016 Sep;54:401--440.

\bibitem{spr10}
{Schwab} J, {Podsiadlowski} P, {Rappaport} S.
\newblock {Further Evidence for the Bimodal Distribution of Neutron-star
  Masses}.
\newblock {\it ApJ}. 2010 Aug;719(1):722--727.

\bibitem{ts99}
{Tauris} TM, {Savonije} GJ.
\newblock {Formation of millisecond pulsars. I. Evolution of low-mass X-ray
  binaries with P\_orb> 2 days}.
\newblock {\it Astron Astrop}. 1999 Oct;350:928--944.

\bibitem{cfr+20}
{Cromartie} HT, {Fonseca} E, {Ransom} SM, {Demorest} PB, {Arzoumanian} Z,
  {Blumer} H, et~al.
\newblock {Relativistic Shapiro delay measurements of an extremely massive
  millisecond pulsar}.
\newblock {\it Nature Astron}. 2020 Jan;4:72--76.

\bibitem{saz78}
{Sazhin} MV.
\newblock {Opportunities for detecting ultralong gravitational waves}.
\newblock {\it Astron Zh}. 1978 Feb;22:36--38.

\bibitem{det79}
{Detweiler} S.
\newblock {Pulsar timing measurements and the search for gravitational waves}.
\newblock {\it ApJ}. 1979 Dec;234:1100--1104.

\bibitem{bb74}
{Balick} B, {Brown} RL.
\newblock {Intense sub-arcsecond structure in the galactic center.}
\newblock {\it ApJ}. 1974 Dec;194:265--270.

\bibitem{sal64}
{Salpeter} EE.
\newblock {Accretion of Interstellar Matter by Massive Objects.}
\newblock {\it ApJ}. 1964 Aug;140:796--800.

\bibitem{xcg+23}
{Xu} H, {Chen} S, {Guo} Y, {Jiang} J, {Wang} B, {Xu} J, et~al.
\newblock {Searching for the Nano-Hertz Stochastic Gravitational Wave
  Background with the Chinese Pulsar Timing Array Data Release I}.
\newblock {\it Res~Astron~Astrop}. 2023 Jul;23(7):075024.

\bibitem{aaa+23a}
{Agazie} G, {Anumarlapudi} A, {Archibald} AM, {Arzoumanian} Z, {Baker} PT,
  {B{\'e}csy} B, et~al.
\newblock {The NANOGrav 15 yr Data Set: Evidence for a Gravitational-wave
  Background}.
\newblock {\it ApJ Lett}. 2023 Jul;951(1):L8.

\bibitem{msr+25}
{Miles} MT, {Shannon} RM, {Reardon} DJ, {Bailes} M, {Champion} DJ, {Geyer} M,
  et~al.
\newblock {The MeerKAT Pulsar Timing Array: the first search for gravitational
  waves with the MeerKAT radio telescope}.
\newblock {\it MNRAS}. 2025 Jan;536(2):1489--1500.

\bibitem{rzs+23}
{Reardon} DJ, {Zic} A, {Shannon} RM, {Hobbs} GB, {Bailes} M, {Di Marco} V,
  et~al.
\newblock {Search for an Isotropic Gravitational-wave Background with the
  Parkes Pulsar Timing Array}.
\newblock {\it ApJ Lett}. 2023 Jul;951(1):L6.

\bibitem{bkh+82}
{Backer} DC, {Kulkarni} SR, {Heiles} C, {Davis} MM, {Goss} WM.
\newblock {A millisecond pulsar}.
\newblock {\it Nature}. 1982 Dec;300(5893):615--618.

\bibitem{sc10}
{Shannon} RM, {Cordes} JM.
\newblock {Assessing the Role of Spin Noise in the Precision Timing of
  Millisecond Pulsars}.
\newblock {\it ApJ}. 2010 Dec;725(2):1607--1619.

\bibitem{ew75}
{Estabrook} FB, {Wahlquist} HD.
\newblock {Response of Doppler spacecraft tracking to gravitational radiation.}
\newblock {\it Gen Rel Grav}. 1975 Oct;6(5):439--447.

\bibitem{lwk+11}
{Lee} KJ, {Wex} N, {Kramer} M, {Stappers} BW, {Bassa} CG, {Janssen} GH, et~al.
\newblock {Gravitational wave astronomy of single sources with a pulsar timing
  array}.
\newblock {\it MNRAS}. 2011 Jul;414(4):3251--3264.

\bibitem{hd83}
{Hellings} RW, {Downs} GS.
\newblock {Upper limits on the isotropic gravitational radiation background
  from pulsar timing analysis.}
\newblock {\t ApJ Lett}. 1983 Feb;265:L39--L42.

\bibitem{pdd+19}
{Perera} BBP, {DeCesar} ME, {Demorest} PB, {Kerr} M, {Lentati} L, {Nice} DJ,
  et~al.
\newblock {The International Pulsar Timing Array: second data release}.
\newblock {\it MNRAS}. 2019 Dec;490(4):4666--4687.

\bibitem{rwr+21}
{Riley} TE, {Watts} AL, {Ray} PS, {Bogdanov} S, {Guillot} S, {Morsink} SM,
  et~al.
\newblock {A NICER View of the Massive Pulsar PSR J0740+6620 Informed by Radio
  Timing and XMM-Newton Spectroscopy}.
\newblock {\it ApJ Lett}. 2021 Sep;918(2):L27.

\bibitem{aaa+23b}
{EPTA Collaboration}, {InPTA Collaboration}, {Antoniadis} J, {Arumugam} P,
  {Arumugam} S, {Babak} S, et~al.
\newblock {The second data release from the European Pulsar Timing Array. III.
  Search for gravitational wave signals}.
\newblock {\it Asrton~Astrop}. 2023 Oct;678:A50.

\bibitem{adg+23}
{Allen} B, {Dhurandhar} S, {Gupta} Y, {McLaughlin} M, {Natarajan} P, {Shannon}
  RM, et~al.
\newblock {The International Pulsar Timing Array checklist for the detection of
  nanohertz gravitational waves}.
\newblock arXiv e-prints. 2023 Apr:arXiv:2304.04767.

\bibitem{gnt+25}
{Grunthal} K, {Nathan} RS, {Thrane} E, {Champion} DJ, {Miles} MT, {Shannon} RM,
  et~al.
\newblock {The MeerKAT Pulsar Timing Array: Maps of the gravitational wave sky
  with the 4.5-yr data release}.
\newblock {\it MNRAS}. 2025 Jan;536(2):1501--1517.

\bibitem{aaa+23c}
{Afzal} A, {Agazie} G, {Anumarlapudi} A, {Archibald} AM, {Arzoumanian} Z,
  {Baker} PT, et~al.
\newblock {The NANOGrav 15 yr Data Set: Search for Signals from New Physics}.
\newblock {\it ApJ Lett}. 2023 Jul;951(1):L11.

\bibitem{wch24}
{Wu} YM, {Chen} ZC, {Huang} QG.
\newblock {Cosmological interpretation for the stochastic signal in pulsar
  timing arrays}.
\newblock Science China Physics, Mechanics, and Astronomy. 2024
  Apr;67(4):240412.

\bibitem{lpr+19}
{Lorimer} D, {Pol} N, {Rajwade} K, {Aggarwal} K, {Agarwal} D, {Strader} J,
  et~al.
\newblock {Radio Pulsar Populations}.
\newblock {\it BAAS}. 2019 May;51(3):261.

\bibitem{wh16}
{Weisberg} JM, {Huang} Y.
\newblock {Relativistic Measurements from Timing the Binary Pulsar PSR
  B1913+16}.
\newblock {\it ApJ}. 2016 Sep;829(1):55.

\bibitem{hf24}
{Hu} H, {Freire} PCC.
\newblock {Measuring the Lense{\textendash}Thirring Orbital Precession and the
  Neutron Star Moment of Inertia with Pulsars}.
\newblock {\it Uni}. 2024 Mar;10(4):160.

\bibitem{mcc14}
{Madison} DR, {Cordes} JM, {Chatterjee} S.
\newblock {Assessing Pulsar Timing Array Sensitivity to Gravitational Wave
  Bursts with Memory}.
\newblock {\it ApJ}. 2014 Jun;788(2):141.

\bibitem{wil98}
{Will} CM.
\newblock {Bounding the mass of the graviton using gravitational-wave
  observations of inspiralling compact binaries}.
\newblock {\it Phys Rev D}. 1998 Feb;57(4):2061--2068.

\end{thebibliography}

\end{document}